\documentclass[aps,prl,twocolumn,floatfix,showpacs]{revtex4}
\usepackage{graphicx,amsmath,amssymb,mathptm}
\begin{document}
\title{Asymmetrically coupled directed percolation systems}
\author{Jae Dong Noh}
\affiliation{Department of Physics, Chungnam National University, Daejeon
305-764, Korea}
\author{Hyunggyu Park}
\affiliation{School of Physics, Korea Institute for Advanced Study, Seoul
130-722, Korea}
\date{\today}

\begin{abstract}
We introduce a dynamical model of coupled directed percolation systems 
with two particle species. The two species $A$ and $B$ are coupled 
asymmetrically in that $A$ particles branch $B$ particles 
whereas $B$ particles prey on $A$ particles.
This model may describe epidemic spreading controlled by reactive 
immunization agents. We study nonequilibrium phase transitions
with focused attention to the multicritical point where both
species undergo the absorbing phase transition simultaneously. 
In one dimension, we find that the inhibitory coupling from $B$ to 
$A$ is irrelevant and the model belongs to the 
unidirectionally coupled directed percolation universality class. 
On the contrary, a mean field analysis predicts that the inhibitory coupling 
is relevant and a new universality appears with a variable dynamic exponent.
Extensive numerical simulations on small-world networks confirm our predictions.
\end{abstract}
\maketitle

It is a challenging problem to establish the concept of the universality
class in nonequilibrium phase transitions. In contrast to equilibrium 
systems, nonequilibrium systems lack a general framework to begin with 
and display too diverse critical phenomena to be classified into a few
universality classes. Recently the absorbing phase transition has been
attracting much theoretical interest since the universality class concept
can be applied. 
It is a phase transition from an inactive (dead) phase into
an active (live) phase, which may occur in systems with 
microscopic trapped (absorbing) states 
which the system, once trapped, cannot excape out of.

Absorbing critical phenomena are categorized into a few universality classes
depending on conservation in dynamics and symmetry between absorbing
states~(see for review Ref.~\cite{Hinrichsen00}). 
The directed percolation~(DP) class is the most well-established
one, which encompasses a vast number of systems~\cite{Janssen81,Grassberger82} .
The contact process~\cite{Harris74} is an archetype of the DP class.
In this model, each lattice site is either empty~($\emptyset$) or 
occupied by a particle~($X$). Each particle may annihilate spontaneously
($X\rightarrow\emptyset$) or branch an offspring to a neighboring site
($X\rightarrow XX$).
The contact process was originally suggested as a model
for epidemic spreading without immunization~\cite{Harris74},
where the particle represents an
infected individual, who may recover spontaneously or infect its healthy 
neighbors. 
The vacuum state with all sites empty 
is the unique absorbing state, which is one of the key ingredients 
in the DP class. 

Recently, coupled DP systems have been studied 
extensively~\cite{Jans,KwonPark,UCDP,UCDP2}. In particular, 
the unidirectionally coupled DP (UCDP) process introduced by
T\"{a}uber {\em et al.}~\cite{UCDP} was found to display a series
of new multicritical phenomena. 
In the UCDP process, there are an
infinite number of particle species $X_k$ with $k=0,1,2,\cdots$
with dynamic rules characterized by $X_k\rightarrow
\emptyset$, $X_k\rightarrow X_k X_k$, and $X_k \rightarrow X_k
X_{k+1}$. The coupling is unidirectional (upward), linear, and excitatory. 
The species $X_0$ is not affected by the others, hence
its absorbing critical phenomena  belong to the DP class.
The other species $X_{k>0}$, however, are dressed by the critical 
fluctuations of particles of species $X_{k'<k}$ at the multicritical point
where all species undergo the absorbing phase transition simultaneously.
It turned out that their critical behaviors are described by the
critical exponents distinct from those of the DP class~\cite{UCDP,UCDP2}.
This novel critical behavior is destroyed and the DP scaling is recovered as soon as 
we turn on the excitatory coupling in the other (downward) direction~\cite{KwonPark,UCDP2}.

It is noteworthy that the UCDP process was proposed for describing 
an interesting roughening transition in an one-dimensional growth 
process~\cite{Alon96}, where the monomer-type restricted-solid-on-solid 
(RSOS) model is considered with the constraint that evaporation is allowed
only at the edges (not terraces). It can be easily shown that
the growth model is mapped to the UCDP process, except that
the RSOS condition generates an extra {\em inhibitory} coupling in the downward
direction, i.e.~the offspring production of $X_k$ is suppressed by the presence 
of $X_{k'}$ for $k'>k$ (see Sec.VA in~\cite{UCDP2}). The coupling should be
at least bilinear, so the simplest process one may consider is
$X_k X_{k+1} \rightarrow X_{k+1}$.
Numerical studies~\cite{Alon96,UCDP2} revealed that the
one-dimensional growth model displays the same critical behavior as the
UCDP, which may imply  the irrelevance of the inhibitory coupling in 
coupled DP processes. 

In this Letter, we study a coupled DP system with two particle species, which
has a linear excitatory coupling in one direction and a bilinear inhibitory coupling
in the other direction. We show that the inhibitory coupling is irrelevant only in low
dimensions (UCDP class) but  becomes relevant in high dimensions, at least where
the mean field theory becomes valid. Our model may also serve as a prototype 
model for epidemic spreading process in the presence of reactive immunization agents,
which will be explained shortly. 

We consider a coupled 
DP system with two particle species $A$ and $B$ on lattices or networks
in general. Specifically we choose the contact process dynamics for
both species in the noninteracting limit; each particle annihilates
spontaneously or branches one offspring of the same kind to a neighboring site.
The two species are coupled {\em asymmetrically} through branching and 
predation processes, that is, an $A$ particle branches a $B$ particle to a 
neighboring site~(excitatory coupling), while a $B$ particle preys on an 
$A$ particle at the same site~(inhibitory coupling). 

The dynamic rules are summarized as follows:
(i) $A~(B) \rightarrow \emptyset$ with probability $p_A~(p_B)$, 
(ii) $A\rightarrow AA$ with probability $(1-p_A)(1-\lambda)$, 
(iii) $B\rightarrow BB$ with probability $1-p_B$, 
(iv) $A \rightarrow AB$ with probability $(1-p_A)\lambda$, and 
(v) $AB \rightarrow B$ with probability $\mu$.
The order parameters for the phase transition are the particle densities,
$\rho_A$ and $\rho_B$.
The model will be referred to as the asymmetrically coupled DP~(ACDP) process.
When $\mu=0$, the coupling becomes unidirectional and the model reduces to
the two-species case of the UCDP process.

In the context of epidemiology, $A$ particles correspond to antigens (or viruses)
and $B$ particles correspond to antibodies (or vaccines). Both antigens and
antibodies can replicate themselves and may die out by themselves 
(processes (i), (ii), and (iii)). In addition, 
an antigen promotes an antibody (iv) and can be destroyed by the antibody (v).

\begin{figure}[t]
\includegraphics*[width=0.7\columnwidth]{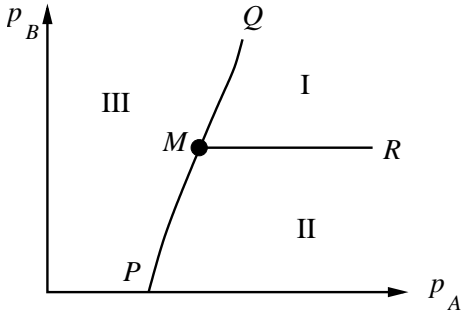}
\caption{A schematic phase diagram of the ACDP process in the $(p_A, p_B)$
plane with fixed $\lambda$ and $\mu$.}
\label{fig1}
\end{figure}

A schematic phase diagram of the ACDP process is shown in 
Fig.~\ref{fig1}. In the phase I
where both annihilation probabilities are large,
both species are inactive with $\rho_A=\rho_B=0$ in the steady state.
When $p_B$ is smaller, the $B$ species becomes active with the $A$ species 
remaining inactive. Hence, in the phase II, $\rho_A=0$ and $\rho_B\neq 0$ in
the steady state. When $p_A$ is small, the $A$ species is active, so
is the $B$ species regardless of the value of $p_B$ due to the process
$A\rightarrow AB$. So, both species are active with $\rho_A\neq 0$ and
$\rho_B\neq 0$ in the phase III. 
Since $\rho_A=0$ throughout the phases I and II, the phase boundary line
$\overline{MR}$ should be flat with $p_B=p_B^c$ which is a constant
independent of $p_A$, $\lambda$, and $\mu$, and is identical to
the critical probability of the uncoupled system.

The nature of the phase transitions is easily
inferred. Across the line $\overline{MR}$, the $B$ species undergoes the
absorbing transition with $\rho_A=0$. Across the line
$\overline{MP}$, the $A$ species undergoes the transition while $\rho_B$
remains finite. Across the line $\overline{MQ}$, both species undergoes
the transition with $\rho_B\propto \rho_A$ since the $B$ particles are
mostly slaved by the $A$ particles through the $A\rightarrow AB$ process for large 
$p_B$ region. Therefore, one may easily expect that the three critical
lines should belong to the DP class. That was confirmed by numerical
simulations, which we do not 
present here. Interesting critical phenomena can be observed along a line
through the multicritical point $M$ from the phase III to the phase I. 
In this Letter, we determine the location of $M$ with high precision and mainly focus on 
density decay dynamics of both species at $M$, which may display 
a novel critical behavior distinct from the DP and also from the UCDP.

First, we investigate the model using the mean field theory which assumes
that local densities of $A$ and $B$ particles are uniform and uncorrelated 
spatially with the mean values $\rho_A$ and $\rho_B$, respectively. 
The mean field rate equations can be written as
\begin{eqnarray}
\dot{\rho}_A &=& a_A \rho_A - \rho_A^2 - \widetilde\mu \rho_A \rho_B,
\nonumber \\
\dot{\rho}_B &=& a_B \rho_B - \rho_B^2 + \widetilde\lambda \rho_A,  
\label{eq:rho_MF}
\end{eqnarray}
where the densities are rescaled to fix the coefficient of the $\rho^2$ 
terms to be unity and  the parameters are functions of the model parameters.
Higher order terms are neglected. 

The stationary state densities are obtained 
by solving Eq.~(\ref{eq:rho_MF}) with 
$\dot{\rho}_A=\dot{\rho}_B=0$ and hence the phase diagram is determined. 
It has the same structure as shown in 
Fig.~\ref{fig1}. The critical lines $\overline{MR}$, $\overline{MQ}$, and
$\overline{MP}$ are given by $a_B=0$ ($a_A\leq0$), $a_A=0~(a_B\leq 0)$, 
and $a_A = \widetilde{\mu} a_B~(a_B\geq0)$, respectively. 
The multicritical point $M$ locates at $(a_A,a_B)=(0,0)$. 

Approaching the multicritical point 
along a straight line $(a_A,a_B)=(\varepsilon,r \varepsilon)$ with fixed $r$ 
inside the phase III, we find that the density
vanishes as $\rho_{A,B} \sim \varepsilon^{\beta_{A,B}}$ with the order
parameter exponents $\beta_A = 2$ and $\beta_B = 1$.
Spatial fluctuations in the order parameter are easily incorporated into the
mean field theory by including diffusion terms $\nabla^2 \rho_{A,B}$ in the
mean field equations. With those, one can show that the characteristic
length scale diverges as $\xi_{A,B} \sim \varepsilon^{-\nu_{A,B}}$
with the correlation length exponent $\nu_A = \nu_B = 1/2$.
The characteristic time scale diverges with the length scale as 
$\tau_{A,B} \sim \xi_{A,B}^{z_{A,B}}$ with the dynamic exponent
$z_A = z_B = 2$.

We note that the critical exponents for the $B$ species are identical to
the mean field DP exponents with
$\beta=1$, $\nu=1/2$, and $z=2$~\cite{Hinrichsen00}. 
On the other hand, the
$A$ species dynamics is dressed by the critical DP fluctuations and falls
into the non-DP class. In the UCDP limit with $\widetilde\mu=0$,
we have the opposite situation; $A$ species falls into the DP class with
$\beta_A=1$ and $B$ species into the non-DP class with
$\beta_B=1/2$~\cite{UCDP}.
From the mean field analysis we conclude that the inhibitory coupling is
relevant and that the ACDP forms an independent universality class distinct from
the UCDP class.

More interesting is the critical density decay dynamics at 
the multicritical point. Putting $a_A=a_B=0$ in
Eq.~(\ref{eq:rho_MF}), after straightforward
calculations, we obtain that the density decays algebraically in time 
as $\rho_A \sim t^{-\alpha_A}$ and $\rho_B \sim t^{-\alpha_B}$
with the exponent $\alpha_B=1$ and the variable exponent
\begin{equation}
\alpha_A = \max\{2,\widetilde{\mu}\} \ .
\end{equation} 
At $\widetilde{\mu}=2$,
logarithmic corrections appear as 
$\rho_A \simeq  t^{-2} (2\widetilde\lambda\ln t)^{-1}$ and
$\rho_B \simeq  t^{-1}(1 + 1/(2\ln t))$. 
In the normal scaling, we expect that the
density decay exponent is given by $\alpha = \beta / (z
\nu)$~\cite{Hinrichsen00}, which does not hold in our model 
for $\widetilde{\mu}>2$.
This is another novel feature of the ACDP universality class. 

The mean field characteristics may be realized on regular lattices
in higher space dimensions than the upper critical dimension and
also on various networks including complete graphs, random networks,
and small-world networks~\cite{Watts98}. Recently, the
small-world network is found to be useful and efficient
in studying mean field critical phenomena~\cite{Kim_Hong} and it
reflects some of network aspects in real biological and social systems.
In this Letter, we adopt the small-world networks to test our mean-field result 
and also to get the information on the upper critical dimension of our model. 

The small-world network is characterized by the total number of sites $N$, 
the average coordination number $2Z$, and
the so-called rewiring probability $p_{r}$. It is constructed as
follows~\cite{Watts98}:
First, $N$ sites are arranged along an one-dimensional ring and each site
is connected up to $Z$ neighboring sites. Second, each bond is rewired
randomly with probability $p_r$. It interpolates the normal
one-dimensional lattice at $p_r=0$ and the random network at $p_r=1$.
In simulations, we choose $Z=10$, $p_r = 0.5$, and $N\leq 2\times 10^6$. 

In networks, the concepts of dimensionality as well as length scale 
are meaningless. 
Botet {\em et al.}~\cite{Botet82} suggested that a correlation volume,
instead of length, $\xi_V$ be a proper scaling variable and the criticality
be described by the correlation volume exponent $\bar{\nu}$ and the
{\em effective} dynamic exponent $\bar{z}$. 
Moreover, they conjectured that the exponents are given by
\begin{equation}\label{eq:conjecture}
\bar\nu = \nu^{MF} d^u \quad \mbox{and} \quad \bar{z} = z^{MF} / d^u 
\end{equation}
where $\nu^{MF}$ and $z^{MF}$ are the mean-field exponents and 
$d^u$ is the upper critical dimension. 
Although the conjecture has not been proved yet, 
it is shown to be valid for some equilibrium systems~\cite{Botet82,Kim_Hong}.
We will use this relation to find the upper critical dimension of our system.

\begin{figure}[t]
\includegraphics*[width=\columnwidth]{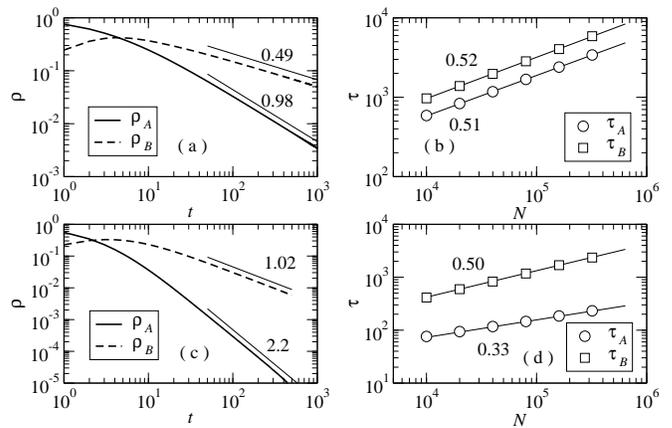}
\caption{(a) Density decay~($N=2\times 10^6$) and (b) characteristic 
time scale at $\mu=0$ and $p_A = 0.3202$. (c) Density decay~($N=2\times 10^6$) 
and (d) characteristic time scale at $\mu=1.0$ and $p_A=0.2902$.}
\label{fig2}
\end{figure}

The phase boundary $\overline{MR}$ is given by a flat line of $p_B =
p_B^c$, which is the critical probability of the uncoupled system. 
We perform Monte Carlo simulations starting with the fully 
occupied configuration of $B$ particles only.
All numerical data shown below are obtained after
average over, at least, several thousands of samples.
As the particle density decays algebraically in time at criticality, 
we can easily 
locate $p_B^c = 0.48505(5)$ where  $\rho_B \sim t^{-\alpha_B}$ 
with $\alpha_B = 0.98(5)$. This confirms that 
the critical line $\overline{MR}$ belongs to the mean-field DP class.

Once $p_B^c$ being found, the multicritical point $M$ can be located 
by examining the decay dynamics of $\rho_A$, starting from the fully 
occupied configuration of $A$ particles, as one varies $p_A$ 
with $p_B=p_B^c$ fixed. For the coupling constants, we fix 
$\lambda=0.5$ and vary $\mu$ . 
We first study the UCDP limit at $\mu=0$. In Fig.~\ref{fig2} (a),
the particle density decay is presented at the multicritical point at
$p_A^c = 0.3202(1)$. The density decay exponents are estimated as
$\alpha_A=0.98(3)$ and $\alpha_B=0.49(2)$ as indicated by straight lines in the
plot. Note that the results are consistent with the mean field results of
the UCDP~\cite{UCDP}. In Fig.~\ref{fig2} (b), we present the scaling of
the characteristic time $\tau$ at which the survival probability becomes 1/2.
It grows algebraically with the total number of sites $N$ with the effective
dynamic exponent
$\bar{z}_A = 0.51(2)$ and $\bar{z}_B = 0.52(2)$ as indicated by straight lines.
Using Eq.~(\ref{eq:conjecture}) and the mean field dynamic exponents 
$z_A=z_B=2$~\cite{UCDP}, 
we obtain that the upper critical dimensionality is given by $d^u=4$ 
for both species, which is consistent with the prediction for
the UCDP class~\cite{UCDP}.

Turning on the inhibitory coupling $(\mu\neq 0)$, we find different scaling
exponents.
In Fig.~\ref{fig2} (c), the density decay is shown 
at the multicritical point $p_A^c=0.2902(2)$ at $\mu=1$.
As predicted in our mean field theory, the estimated decay exponent 
$\alpha_B=1.02(3)$ 
is consistent with the mean-field DP value and $\alpha_A=2.20(5)$ represents 
the non-DP class. 
We also estimate the effective dynamic exponents
$\bar{z}_A=0.33(1)\simeq 1/3$ and $\bar{z}_B=0.50(1)$ in  Fig.~\ref{fig2} (d).
Since $z^{MF}=2$ for both species, Eq.~(\ref{eq:conjecture}) 
suggests that the two species have different upper critical
dimensions, i.e.~$d_A^u = 6$ and $d_B^u = 4$.
It is remarkable to see that the two interacting species begin to show the
mean field critical behaviors at different dimensionality. 
A simple power counting for coupled Langevin equations obtained 
from Eq.~(\ref{eq:rho_MF}) by adding diffusion terms and DP-type 
multiplicative noises reproduces the above results, but 
more efforts  are needed to understand this peculiar property both
analytically and numerically. 

\begin{figure}[t]
\includegraphics*[width=\columnwidth]{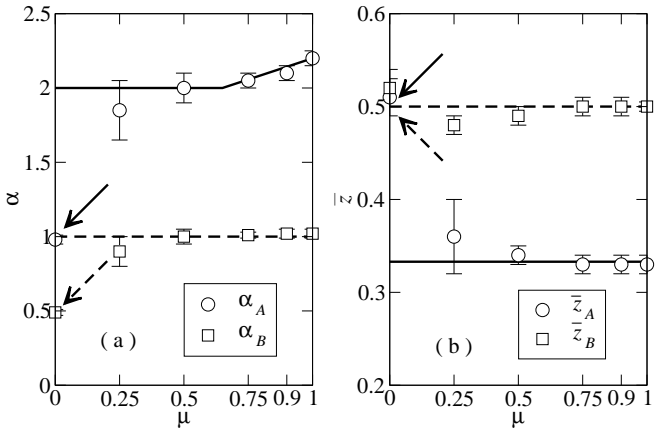}
\caption{(a) The density decay exponent and 
(b) the effective dynamic exponents at the multicritical points.
The solid~(dashed) lines are guides to eyes and indicate the mean field
results for the $A~(B)$ species of the ACDP process.
The solid~(dashed) arrows point toward the corresponding exponent values of 
the mean field UCDP class.
}\label{fig3}
\end{figure}

We performed the similar analysis for various values of $\mu$. The results
for the decay exponents and the effective dynamic exponents are presented in
Fig.~\ref{fig3}. The numerical results are in good agreement with the mean
field results. Near $\mu=0$, e.g., at $\mu=0.25$,  
the exponents seem to deviate from the mean field values. These are
attributed to crossovers due to the nearby UCDP fixed point.
The mean field theory and the numerical results suggest that the ACDP 
process with $\mu\neq 0$ form a distinct universality class different
from the UCDP class.

\begin{figure}[t]
\includegraphics*[width=\columnwidth]{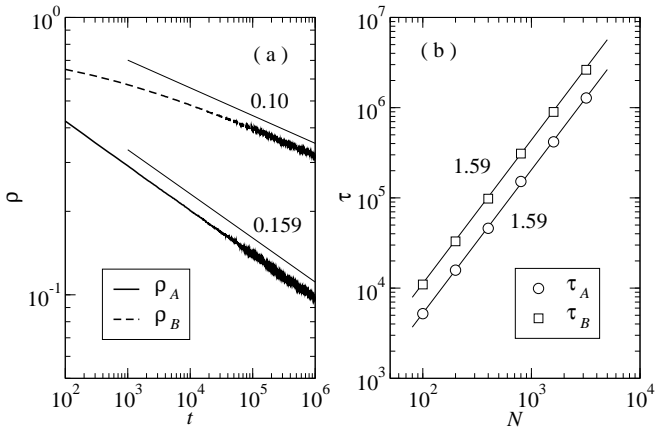}
\caption{(a) Density decay at the multicritical point of
the ACDP model with $\lambda = 0.2$ and $\mu = 0.2$ on the one-dimensional
lattice with $N=10^5$ sites. (b) Characteristic time scales at the
multicritical point.}
\label{fig4}
\end{figure}

The previous study poses a question on the relevance of the inhibitory
coupling at lower dimensions~\cite{UCDP}. We check the multicritical behavior of
our model  on the one-dimensional lattice. 
We fix the coupling constants $\lambda = 0.2$ and $\mu=0.2$.
Numerical simulations show that the multicritical point is 
located at $(p_A, p_B)=(0.14627(2),0.23267)$. 
The density decay and the characteristic time scales are presented in
Fig.~\ref{fig3}. The decay exponents are estimated as
$\alpha_A = 0.159(5)$ and $\alpha_B = 0.10(2)$, which are
consistent with the results for the UCDP class in one dimension~\cite{UCDP}.
The dynamic exponents are esimated as $z_A \simeq z_B\simeq 1.59(2)$, 
which are consistent with the DP value. 
These results suggest that the ACDP process belongs to the UCDP class in one
dimension. That is, the inhibitory coupling is irrelevant in one dimension in
the contrary to the higher dimensional cases.

In summary, we have introduced the ACDP process 
of two particle species and investigated its multicritical behaviors.
In the mean field theory, we show that the inhibitory coupling is
relevant and the ACDP process forms the different universality class
distinct from the UCDP process: The $B$ species displays the DP type critical
behavior while the $A$ species the dressed non-DP type critical behavior.
Furthermore, the critical dynamics of the $A$ species is anomalous with the
variable critical exponent. The mean field results are confirmed by the
numerical simulation studies of the ACDP on the small-world networks.
Using the conjecture in Eq.~(\ref{eq:conjecture}), we also find that
the upper critical dimensionality for each species is given by $d_A^u = 6$
and $d_B^u=4$. On the other hand, the numerical simulation studies in
one dimension show that the inhibitory coupling is irrelevant and the ACDP
process belongs to the UCDP class: The $A$ species displays the DP type
critical behavior while the $B$ species the dressed non-DP type critical
behavior.

For future works, we suggest that it would be interesting to study the model
in higher dimensional lattices to find the dimensionality at which the
inhibitory coupling becomes relevant and to cofirm the upper critical
dimensionalities. The multi-species generalization would 
be also interesting~\cite{multi}. As an application, 
the phase transition of the ACDP process on general complex networks 
may yield richer critical dynamics. 

This work was supported by Korea Research Foundation 
Grant (KRF-2004-041-C00139).


\end{document}